\documentclass{elsart}

\usepackage{graphicx}

\usepackage{amssymb}

\begin{document}

\begin{frontmatter}
\journal{Astroparticle Physics}



\title{On the Evidence for Clustering
in the Arrival Directions of AGASA's Ultrahigh Energy Cosmic Rays}


\author{Chad B. Finley,\corauthref{cor1}}
\author{Stefan Westerhoff}

\address{Columbia University, Department of Physics, 
New York, NY, USA}
\corauth[cor1]{corresponding author: finley@phys.columbia.edu}

\begin{abstract}
Previous analyses of cosmic rays above $4\cdot 10^{19}$\,eV
observed by the AGASA experiment have suggested that
their arrival directions may be clustered.  However, estimates of the
chance probability of this clustering signal vary from
$10^{-2}$ to $10^{-6}$ and beyond.
It is essential that the strength of this evidence be well
understood in order to compare it with anisotropy studies in
other cosmic ray experiments.  We apply two methods
for extracting a meaningful significance from this data set:
one can scan for the cuts which optimize
the clustering signal, using simulations to determine the
appropriate statistical penalty for the scan.  This analysis
finds a chance probability of about $0.3\%$.  Alternatively,
one can optimize the cuts with a first set of data, and then
apply them to the remaining data directly without statistical penalty.
One can extend the statistical power of this test by considering
cross-correlation between the initial data and the remaining data,
as long as the initial clustering signal is not included.
While the scan is more useful in general,
in the present case only splitting the data set offers an unbiased test of the 
clustering hypothesis.  Using this test 
we find that the AGASA data is consistent
at the $8\%$ level
with the null hypothesis of isotropically distributed arrival directions.
\end{abstract}


\end{frontmatter}

\section{Introduction}

The study of arrival directions of cosmic rays above $10^{19}$\,eV 
(ultrahigh energy cosmic rays) is one of the most promising ways 
to gain insight into the origin of these particles.  While 
a number of experiments have shown that the distribution of arrival 
directions is remarkably isotropic, 
evidence for small-angle clustering has been claimed, 
most notably by the AGASA~\cite{chiba92} (Akeno Giant Air Shower 
Array) cosmic ray 
experiment~\cite{agasa96,agasa99,agasa00,agasa01,agasa03,agasaweb}.  
This clustering signal, if confirmed, would give strong support to 
the idea that cosmic rays originate from compact 
sources~\cite{tinyakov01}.

The focus on small-angle anisotropies among the very highest
energy events is well-motivated: if the cosmic ray particles are charged, 
then the highest 
energy ones are expected to suffer
the smallest deflections while traversing Galactic 
and extragalactic magnetic fields. Their arrival directions
are therefore the most
likely ones to point back toward sources.

A search for clustering among the highest energy cosmic ray events must
make choices for the minimum energy $E_c$ which defines the data set
and the maximum angular separation $\theta_c$ which defines a pair.
On the one hand, choosing a higher energy 
threshold $E_c$ should reduce deflections and allow clusters to show
up within smaller angular separations $\theta_c$.  This 
holds especially for a detector such as AGASA in which
the angular resolution improves at higher energies.
On the other hand, as a function of energy $E$ the 
cosmic ray flux drops faster than $E^{-2}$, so the statistical 
power of the available data quickly weakens with higher energy thresholds.

For a precise model of cosmic ray source distributions and Galactic and
extragalactic magnetic fields, these competing forces 
would imply optimal choices for $E_c$ and $\theta_c$ to maximize
the clustering signal.  At present, however, not nearly enough is
known about any of these to make {\it a priori} choices useful.
Instead, what is done explicitly or implicitly is to scan over a 
range of values for $E_c$ and
$\theta_c$, and identify the values which maximize the clustering
signal.  In this case, the final significance of the result must
include a penalty factor for the {\it a posteriori} cuts arrived at
by scanning.  

It is the various ways of handling this penalty 
factor---or, in some cases, the failure to include it at all---which 
has led to a wide range of significances attached to 
the AGASA clustering signal.

Evaluating this significance rigorously is crucial for understanding
the anisotropy results of new cosmic ray experiments.
The world data set of detected cosmic ray particles above 
$10^{19}$\,eV is currently dominated by the events observed 
by AGASA, which has been operated continuously since 1990.  In the near 
future, a statistically independent data set from the currently 
operating HiRes~\cite{hiresweb} (High Resolution Fly's Eye) air fluorescence 
detector will become available, and in the more distant future,
the Pierre Auger Array~\cite{augerweb} is expected to produce
an even larger data set of ultrahigh energy cosmic rays.

To compare the anisotropy results of AGASA and these new experiments,
the strength of the AGASA clustering signal must be well understood.
We apply two methods for evaluating its significance.
The first is a general method, applicable
to upcoming searches by other experiments as well; the second is
a specific test of the clustering hypothesis which is meaningful only
in the context of the AGASA data set.

First, we propose that a clustering signal among the highest energy events
can be best evaluated by scanning 
{\it simultaneously} over energy thresholds and angular 
separations to find the values for 
$E_c$ and $\theta_c$ which optimize the signal.
The chance probability of the signal is determined
by counting the number of simulated data sets which yield a stronger
signal under an identical scan.  With this procedure, the statistical 
significance is determined without treating {\it a posteriori} 
cuts as {\it a priori} ones.

As will be shown later, however, a bias remains in the case of the 
AGASA data set
due to the inclusion of the events that led to the
clustering hypothesis in the first place.
One can avoid this bias by removing the early data and
only scanning over the events which
have been detected since the original claim.
Alternatively, one can test the AGASA clustering hypothesis
by applying the original cuts to the newer events directly.
Since the cuts are now {\it a priori}, this test 
requires no statistical penalty.
It has the virtues of being simple and rigorously unbiased.

The paper is organized as follows.  In Section 2, we summarize
previous estimates of the significance of the AGASA clustering
signal and motivate the need for a re-analysis.  
In Section 3, we motivate and describe the 
autocorrelation scanning 
technique applied in our analysis.  In Section 4, we perform
this scan on the published AGASA data set and compare the result
with previous estimates of the significance.  In Section 5,
we address the bias introduced by using the whole data set, 
and perform an analysis using only the unbiased event set.
To increase the statistical power of this test, we also include
the effect of cross-correlation between the original and the unbiased
event sets.  We present our conclusions in Section 6.

\section{Small Scale Clustering in AGASA Cosmic Ray Data}

The AGASA experiment reported possible clustering in the arrival 
directions of ultrahigh energy cosmic rays as early as 
1996~\cite{agasa96}, and has updated this data sample and analysis
in several publications~\cite{agasa99,agasa00,agasa01,agasa03}.
The first report of clustering in 1996 identified three pairs of 
events with angular separation less than 
$2.5^{\circ}$ among the 36 events with energies above 
$4\cdot 10^{19}$\,eV.  The corresponding chance probability
was found to be 2.9\%.  It was noted that the angular separation
of $2.5^{\circ}$ is ``nearly consistent with the measurement error 
($\sqrt{2}\cdot 1.6^{\circ}$)''~\cite{agasa96}.  The 
minimum energy of $4\cdot 10^{19}$\,eV was justified under the assumption 
that the Greisen-Zatsepin-Kuzmin (GZK) cutoff~\cite{greisen66,zatsepin66} 
should lead to an accumulation of events around $4\cdot 10^{19}$\,eV,
and therefore that events above this energy may point back to nearby
sources.  The values for $E_c$ and $\theta_c$ identified in this
report set the stage for all analyses which followed.  

In 1999, a new
publication by AGASA~\cite{agasa99} identified a stronger clustering
signal using these cuts with an enlarged data set now containing 47 events.
The following year, AGASA published an updated list with 57 
events above $4\cdot 10^{19}$\,eV observed
through May, 2000~\cite{agasa00}. 
There is also an additional event below $4\cdot 10^{19}$\,eV
which was added to the list because it forms another 
doublet.\footnote{
This is an unfortunate source of confusion.
Like many authors, we do not include this extra event in our analysis because
it is not clear how many additional events there are between it and
$4\cdot 10^{19}$\,eV.  However, it is sometimes included in other
authors' analyses to which we refer.
}
Not counting the extra event, 
there are four doublets and one triplet in this set. 

This set was analyzed by Tinyakov and 
Tkachev~\cite{tinyakov01}, who calculated the 
chance probability 
as a function of the threshold energy $E_c$ of the data set, 
while keeping the angular bin size constant at 
$2.5^{\circ}$.  The lowest probability was found to be less than 
$10^{-4}$ with $E_c=4.8\cdot 10^{19}$\,eV.  Since this probability 
was obtained by scanning over energies, it does not reflect the 
true significance of the clustering signal.  To estimate the
correct chance probability, the authors numerically calculated a 
correction factor by generating $10^{3}$ random sets of events 
which were then subjected to the same scanning in $E_c$.  27 (3) 
random samples had a probability of less than $10^{-2}$ 
($10^{-3}$), and the authors concluded that the correction 
factor was of order 3.  The final chance probability was given as 
$3\cdot 10^{-4}$, considerably lower than the chance probability
reported by the AGASA collaboration in the original 
publications~\cite{agasa96,agasa99}.

A similar scan was then performed in the size of the angular bin,
{\it i.e.} the maximum angular distance between events 
that defines a cluster.  The probability shows a minimum at
$2.5^{\circ}$, but since this was interpreted as the angular
resolution of the experiment, no correction factor was applied
to the final chance probability.

In~\cite{agasa01} in 2001, 
the AGASA group applied this scanning technique again 
to a data set which was now reported to include 59 events above
$4\cdot 10^{19}$\,eV---essentially the same data set as the one
published in 2000~\cite{agasa00}, though it is unclear whether the one 
event below the energy cutoff was kept, or whether one 
or two new events were added.  Five doublets 
and one triplet were reported in the sample.  A scan over
angular separations was again performed, showing the peak
at $2.5^{\circ}$.  Performing a scan over energies, the significance
of the clustering above $4\cdot 10^{19}$\,eV was said to
be $4.6\, \sigma$, and above $4.5\cdot 10^{19}$\,eV it was
said to be in excess of $5\, \sigma$.  No statistical 
penalties were applied for either the energy or angular
separation scan.  

The most recently published study by AGASA~\cite{agasa03} 
in 2003 recapitulates much of the above analysis.
The same 59 events are analyzed, 
though the AGASA experiment has continued to observe
ultrahigh energy cosmic rays and has reported 
72 events above
$4\cdot 10^{19}$\,eV seen through the end of July, 2002~\cite{agasaweb}.
Forgoing a scan over energies, the chance probability for
all of the clusters (one triplet + five doublets = eight
pairs) in the total set of 59 events is simply reported to
be less than $10^{-4}$.    

In evaluating the significance of the clustering signal,
it is essential to determine whether the 
original choices of $E_c=4\cdot 10^{19}$\,eV and
$\theta_c=2.5^{\circ}$ were {\it a priori}.

We consider what would
have been required to formulate such an {\it a priori} hypothesis.
In the case of the angular resolution of the experiment,
Monte Carlo simulations can be used to determine the optimal
angular size for a cluster search.  Such a study needs to
take into account that the angular resolution for a ground
array depends on a variety of factors.  For the AGASA detector,
the angular error continues to shrink with increasing energy.
At $10^{20}$\,eV, AGASA reports $\theta_{err} <
1.2^{\circ}$~\cite{agasa99}.  Eight of the 57 events in the
data set are in fact above this energy.

In addition, the angular resolution of ground arrays depends
on the alignment of the shower with the detector array.
In general the errors will be asymmetric.  In~\cite{agasa99},
AGASA reports on the accuracy of the arrival direction
determination by showing the opening angle distribution 
between simulated and reconstructed arrival directions.
The ratios of the
68\,\% and 90\,\% opening angles shown in~\cite{agasa99} are
clearly not those of a circular, two-dimensional Gaussian
distribution.  These complications mean that standard
formulae do not apply and the optimal angle for a cluster
search cannot be stated simply as $\sqrt{2}\cdot 1.6^{\circ}$.

Furthermore, the search angle which optimizes the clustering signal will
also depend on the expected background of chance clusters.
For small data sets, the chance occurrence of a pair is small,
and the signal to noise ratio can be optimized with a larger
separation angle in the search~\cite{alexandreas93}.

In summary, a clustering search which is {\it a priori} should begin
by first using a Monte Carlo simulation to identify the optimal
opening angle size.  The first clustering paper~\cite{agasa96}
gives no indication that such a search program was
undertaken, nor does it claim that $2.5^{\circ}$ is an {\it a
priori} choice.  It merely observes that the clustering signal is
strongest for $\theta_c=2.5^{\circ}$, a value which coincides
to some extent 
(but only approximately, as $\sqrt{2}\cdot 1.6^{\circ}=2.26^{\circ}$)
with the angular resolution of the
experiment around the given energy.  This makes the value interesting,
but not {\it a priori}.

We motivated the scan over threshold energies $E_c$ in Section 1 
on physical grounds.  Here, we note that
the original paper~\cite{agasa96} does not restrict the analysis
to $4\cdot 10^{19}$\,eV, but mentions at least two other
energy thresholds that were looked at as well ($5\cdot 10^{19}$\,eV
and $6.3\cdot 10^{19}$\,eV).  This approach is certainly
valid, for the reasons we mentioned earlier.  However, 
it does not constitute an {\it a priori} search program,
which demands a choice for $E_c$ and $\theta_c$ prior to
examination of the data.  Because the values of $E_c$ and $\theta_c$
are determined by examining the data, a calculation of
the {\it a priori} probability
does not represent the true significance of the observation.
Either the cuts must be tested with independent data, or the statistical
penalty must be evaluated and included in the calculation of the
chance probability.

\section{Scanning and Evaluation of Chance Probability}

The competition between magnetic deflections and statistical
power described earlier offers one motivation
for scanning over small angles among the
highest energy events to locate a signal.
Scanning is especially well motivated in the case of AGASA, where the 
energy dependence of the angular resolution means that clustering can be
better resolved at higher energies.

In our analysis, we perform a scan simultaneously over energy thresholds
and maximum separation angles to find the $E_c$ and $\theta_c$ which 
maximize the clustering signal, and then we perform identical 
scans over simulated data sets to evaluate the true significance. 

In practice, rather than scanning directly over energy thresholds, 
we rank the events by energy and scan over events $N$. 
That is, for each value of $N$ and $\theta$, we restrict
ourselves to the $N$ highest-energy events, and count the number of
pairs $n_p$ separated by less than $\theta$. 
Just as in the usual two-point correlation function, multiplets are
counted by the individual number of pairs which they contain.  A triplet
of events, for example, will be counted as two or three pairs, depending
on the individual separations of the three events.

Prior to scanning the data, we generate a large number ($10^7$) of
simulated data sets with the same exposure as the detector, 
and use them to generate a table of values $P_{mc}$, where 
$P_{mc}(N,\theta,n)$ is the fraction of sets in which the first 
$N$ events contain exactly $n$ pairs separated by less than 
$\theta$.  

For each $(N, \theta)$, the number of pairs $n_p$ is counted in 
the data, and the probability $P_{data}$ for observing $n_p$ 
{\it or more pairs} at $(N, \theta)$ is calculated as:

\begin{equation} 
P_{data}(N, \theta) =
   \sum_{n=n_p}^{\infty} P_{mc}(N,\theta, n) = 
          1 - \sum_{n=0}^{n_p-1} P_{mc}(N, \theta, n). 
\end{equation}

For some combination $(N_c, \theta_c)$, $P_{data}$ has a minimum: 
$P_{min} = P_{data}(N_c, \theta_c)$.  This identifies the location 
in the scan of the strongest potential clustering signal.  To 
assess the true significance of this signal, we perform the same 
scan over $n_{mc}$ Monte Carlo data sets, identifying the minimum 
probability $P_{min}^{i}=P^{i}(\theta^i_c,N^i_c)$ for each trial 
and counting the number of trials $n_{mc}^{\ast}$ for which 
$P_{min}^{i}\le P_{min}$.

The chance probability of observing $P_{min}$ in the scan is finally
evaluated as:

\begin{equation}
P_{chance} = \frac{n_{mc}^{\ast}}{n_{mc}}.
\end{equation}

This scanning technique is essentially an auto-correlation analysis 
in which the angular size of the first bin and the energy threshold 
of the data set are varied to maximize the signal, and the final 
significance includes the correction factor for the scan over both 
variables.

\section{Autocorrelation Scan of the AGASA Data Set}

We perform this scan on the published AGASA data above 
$4\cdot 10^{19}$\,eV, which consists of 57 events~\cite{agasa00}.  
To generate Monte Carlo events for determining the probabilities, 
we follow~\cite{tinyakov01} in using a zenith angle ($\theta_z$) 
distribution $dn \propto \cos \theta_z \sin \theta_z d\theta_z$, 
corresponding to geometric acceptance of isotropically distributed 
cosmic ray arrival directions.  We use the same $\theta_z<45^{\circ}$ 
cut as employed by AGASA, and assign uniformly random arrival times, 
corresponding to the uniform exposure of AGASA in right 
ascension~\cite{agasa00,agasa01}.  We scan over angular separations
from $0^{\circ}$ to $5^{\circ}$ in increments of $0.1^{\circ}$.  The 
results of the scan are shown in Figure 1.

\begin{figure}[t]
\includegraphics[width=1.\textwidth]{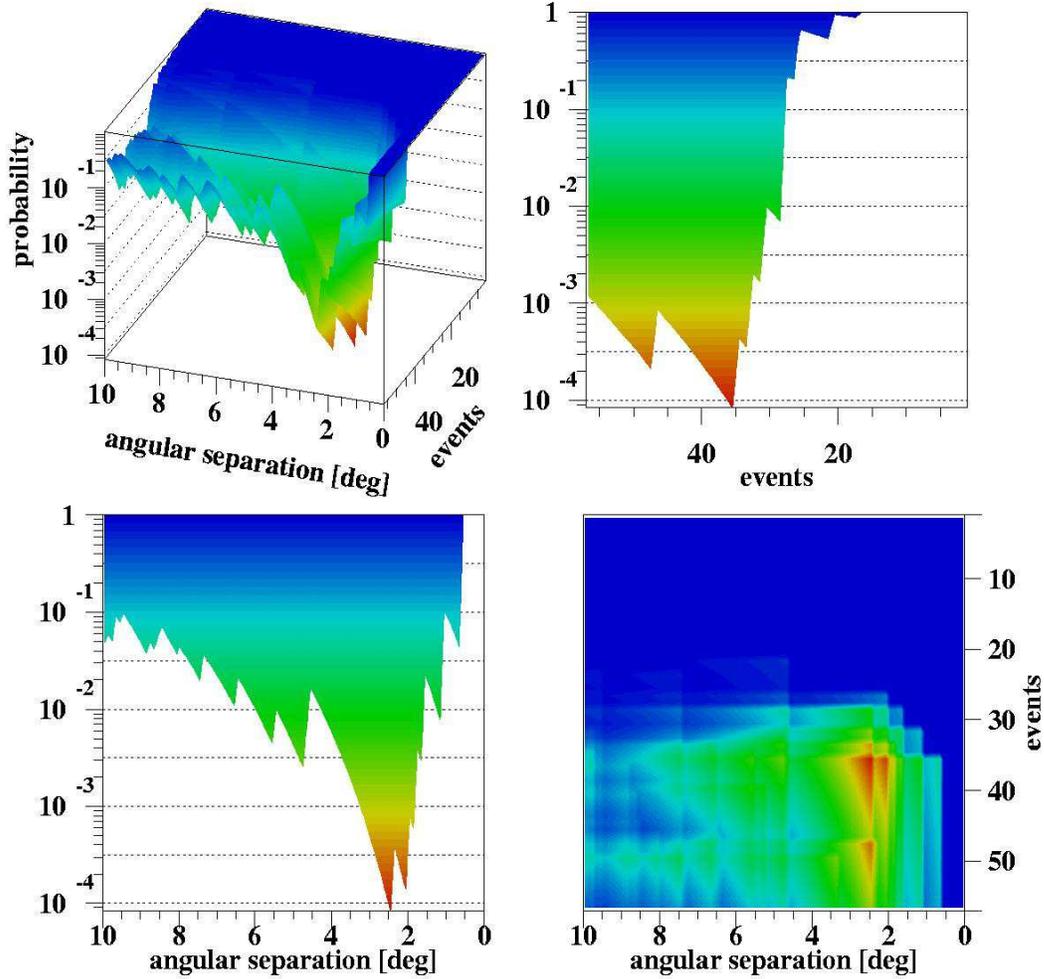}
\caption{\it Scan of AGASA events above $4\cdot 10^{19}\,\mathrm{eV}$, 
shown in four different views.  
$P_{min}=8.39\cdot 10^{-5}$ and $P_{chance}=0.3\%$ for the
clustering signal at $N_c=36$, 
$\theta_c=2.5^{\circ}$ with $n_p=6$ pairs.  
{\rm (}$N_c$ corresponds to an energy 
threshold of $4.89\cdot 10^{19}\,\mathrm{eV}$.{\rm )}}
\end{figure}

The strongest clustering signal is contained within the $N_c=36$ 
highest-energy events, where there are $n_p=6$ pairs separated by 
less than $\theta_c=2.5^{\circ}$.  (The energy threshold corresponding 
to this subset is $4.89\cdot 10^{19}$~eV.)  At this spot 
$P_{data}=P_{min}=8.4\cdot 10^{-5}$, that is, 839 out of $10^7$ MC 
data sets had the same or greater number of pairs at the same values 
for $N$ and $\theta$.  This value for $P_{min}$ (not $P_{chance}$) 
is essentially the same as the $10^{-4}$ probability found by Tinyakov 
and Tkachev~\cite{tinyakov01} for the same energy threshold and angular 
separation.

To evaluate the significance of this result, we perform the same scan 
over simulated AGASA data sets and count how many simulated sets have 
$P_{min}^{MC} \le P_{min}^{data}$.  We find that 3475 out of $10^6$ 
simulated sets meet this condition, implying a chance probability 
of $0.3\%$.  Figure 2 illustrates 
how $P_{chance}$ varies as a function of $P_{min}$ 
for the simulated AGASA sets.

\begin{figure}[t]
\begin{center}
\includegraphics[width=.85\textwidth]{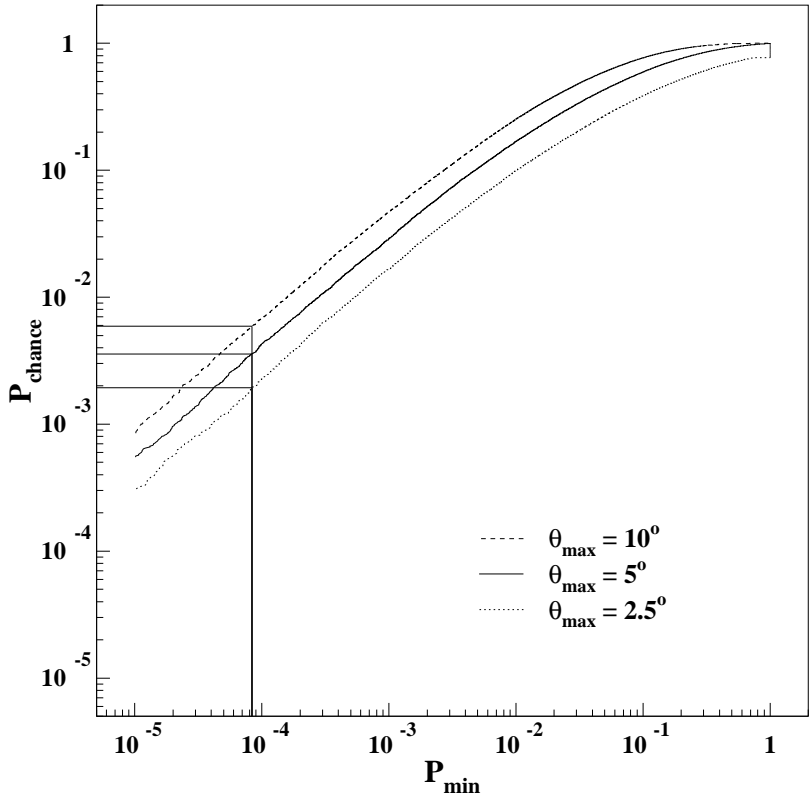}
\end{center}
\caption{\it $P_{chance}$ as a function of $P_{\min}$ for the AGASA
data, with $N_{max}=57$ and three different values of the scan parameter 
$\theta_{max}$.}
\end{figure}

In performing the data scan and the simulated scans, it is necessary 
to choose four parameters which can affect the final result: $N_{max}$,
the total number of events included in the scan; $\theta_{min}$ 
and $\theta_{max}$, 
the angular extent of the scan; and $\Delta \theta$, the size of the 
angular binning.  In Table 1 we show a range of values for these 
parameters and the effect on the final value of $P_{chance}$.  
We motivate our choices for each of the parameters as 
follows:

In choosing the extent of the scan in $N_{max}$ and $\theta_{max}$,
it is clear that the significance would be biased if one scanned out 
precisely to the maximum clustering signal and no further.  
An investigator who reports a clustering signal in a scan can reasonably
be expected to have scanned out at least twice as far in search
of an even stronger signal; hence a reasonable 
estimate of $P_{chance}$ should extend $N_{max}$ to 
$\sim 2\cdot N_c$, and further, if $N_c$ is very small.  The same can 
be said for $\theta_{max}$ with respect to $\theta_c$.  As for $\theta_{min}$,
it should be no larger than the best attainable angular resolution; in
the present case, one could choose $0^{\circ}$ or $1^{\circ}$ with 
little effect on the final probability.
Finally, it can
be seen in Table 1 that reducing the angular bin size $\Delta \theta$
also has a negligible effect at small scales.

\begin{table}
\begin{center}
\begin{tabular}{|c|c|c|c||c|}
\hline
$N_{max}$  & $\theta_{min}$  &  $\theta_{max}$  & $\Delta \theta$  &  $P_{chance}$  \\
\hline
\hline
57 & $0^{\circ}$ & $5^{\circ}$ & $0.1^{\circ}$ &  $3.48\cdot 10^{-3}$  \\
\hline
\hline
{\bf 36}  &   $0^{\circ}$  &   $5^{\circ}$  &  $0.1^{\circ}$ &  $2.40\cdot 10^{-3}$  \\
{\bf 100} &   $0^{\circ}$  &   $5^{\circ}$  &  $0.1^{\circ}$ &  $5.63\cdot 10^{-3}$  \\
\hline
57  &   ${\bf 1^{\circ}}$  &   $5^{\circ}$  &  $0.1^{\circ}$ &  $2.96\cdot 10^{-3}$  \\
57  &   ${\bf 2.5^{\circ}}$ &  $5^{\circ}$  &  $0.1^{\circ}$ &  $2.05\cdot 10^{-3}$  \\
\hline
57  &   $0^{\circ}$  & ${\bf 2.5^{\circ}}$  &  $0.1^{\circ}$ &  $2.00\cdot 10^{-3}$  \\
57  &   $0^{\circ}$  &  ${\bf 10^{\circ}}$  &  $0.1^{\circ}$ &  $5.77\cdot 10^{-3}$  \\
\hline 
57  &   $0^{\circ}$  &   $5^{\circ}$  &  ${\bf 0.5^{\circ}}$ &  $2.31\cdot 10^{-3}$  \\
57  &   $0^{\circ}$  &   $5^{\circ}$  & ${\bf 0.02^{\circ}}$ &  $4.05\cdot 10^{-3}$  \\
\hline
\end{tabular}
\vskip 1cm
\caption
{\it The effect on $P_{chance}$ due to variations in the scan 
parameters.  In each case, $P_{chance}$ was determined using 
$10^{6}$ Monte Carlo data sets.  The top row lists 
the values used in the text.  The parameters which are varied 
are indicated in bold.}
\label{scan_range}
\end{center}
\end{table}

We note that over each of the ranges shown in Table 1---a factor 
of three in event number, a factor of 16 in angular area, and a factor 
of 25 in angular binning---the chance probability
 remains within $2\cdot 10^{-3}$ to $6\cdot 10^{-3}$.  
Therefore our result does not depend sensitively
on these scanning parameters.

The value of $0.3\%$ we calculate for $P_{chance}$ is 10 times larger 
than that calculated by Tinyakov and Tkachev in~\cite{tinyakov01}.  
Although they use an angular scan to demonstrate that the separation 
angle $2.5^{\circ}$ maximizes the signal, they nevertheless treat
the choice as an {\it a priori} one and make no correction for it.

In~\cite{agasa01}, the AGASA collaboration analyzes the same data 
set and finds that at $4\cdot 10^{19}$\,eV the significance of the 
clustering signal is $4.6\, \sigma$, and that at a slightly higher
energy threshold it is ``$5\, \sigma$ or more''.  
These results imply chance probabilities of 
$4.2\cdot 10^{-6}$ and $5.7\cdot 10^{-7}$, respectively---three 
to four orders of magnitude lower than the probability we have
presented here.  This overestimation of the significance of
the clustering signal arises in part from the application of Gaussian
statistics to a non-Gaussian distribution: these significances are 
obtained by measuring the excess clustering signal in units of standard
deviations, $(N_{obs}-N_{exp})/{\Delta N_{exp}}$, when in fact 
this distribution is not Gaussian for the small numbers of clusters
observed.  Having cited Tinyakov and Tkachev 
\cite{tinyakov01} and made use of their technique, the authors 
ignore their warning on exactly this point.  Furthermore, they ignore
the statistical penalty involved in scanning over energy thresholds, and
they do not consider a penalty for the choice of angular separations.

\section{Unbiased Test of AGASA Clustering Hypothesis}

A more rigorous statistical test of the clustering hypothesis can be 
performed by isolating the data which led to the original cuts of 
$4\cdot 10^{19}$\,eV and $2.5^{\circ}$.  Since these values
are justified {\it a posteriori}
in conjunction with the observation in 1996 that they lead to 
a clustering signal~\cite{agasa96}, they can only be treated
as {\it a priori} for a data set independent of the one which was used
to derive them.  We can do this by dividing
the AGASA data into an ``original data set'' comprising the
events observed through October 1995 which formed the basis of the 
original clustering claim, and a ``new data set'' comprising the events
which have been observed since then.  Using the list of events published
in 2000~\cite{agasa00}, there are 30 events
in the original set and 27 in the new one.  
\footnote{In \cite{agasa96} (1996), 
the original data set is said to 
contain 36 events above $4\cdot 10^{19}$\,eV.  
However, the lists \cite{agasa99,agasa00} published in 1999 and 2000
contain only 30 events
during this same time period, due to a reevaluation of the energies  
(according to Uchihori {\it et al.}~\cite{uchihori00}).  
The three original clusters are present in all sets.}

Because the new data set is independent, we can test the 
original clustering hypothesis
directly without the need for any statistical penalties.
We simply count the number of pairs of events using $E_c=4\cdot 10^{19}$\,eV 
and $\theta_c = 2.5^{\circ}$, and we find one pair.
The chance probability for one or more pairs among 27 events is 28\%.

We can investigate whether there is a better choice of 
$E_c$ and $\theta_c$ for the new data set by performing an autocorrelation
scan.  Figure 3 shows the results of scanning over
both the old and new data sets separately.  
The strongest clustering signal in the original set 
has a chance probability $P_{ch} = 4.4\%$ and occurs
for $\theta_c = 2.4^{\circ}$ and $E_c = 4.35 \cdot 10^{19}$~eV
(with 3 pairs among the 26 highest energy
events, and a minimum probability $P_{min} = 0.33\%$ in this bin).
This confirms that the cuts originally selected in 1996 were
nearly optimal for that data set.
However, when the new events are scanned,
there is no hint of clustering
at the $2.5^{\circ}$ scale or any other angular separation.  The 
``strongest'' clustering signal occurs at $\theta_c=4.7^{\circ}$ with
$P_{ch} = 27\%$.

\begin{figure}[t]
\includegraphics[width=1.\textwidth]{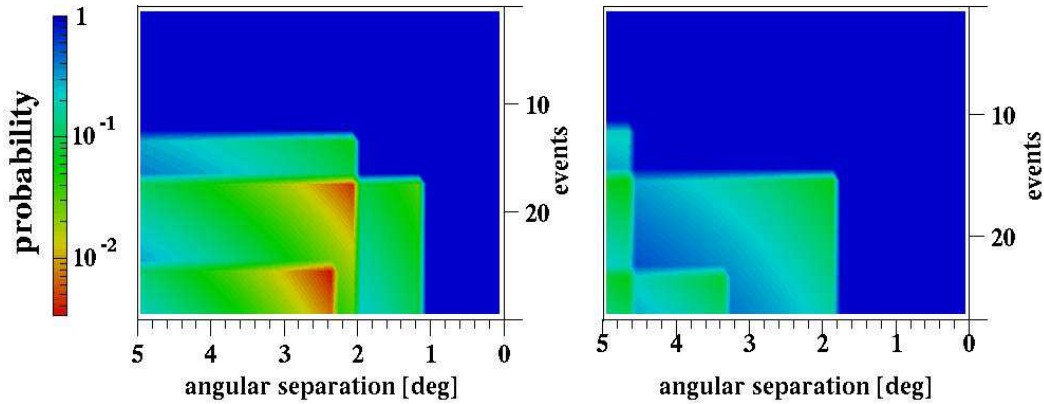}
\caption{\it Autocorrelation scans for the
``original'' (left) and ``new'' (right) AGASA data sets,
using October 31, 1995 as the dividing point.  The chance probability
of the strongest clustering signal in the original data set is 4.4\% (at
$\theta_c = 2.4^{\circ}$, $N_c=26$, $E_c=4.35\cdot 10^{19}$~eV, with
$P_{min}=0.33\%$).  
In the new data set, 
the strongest clustering signal has $P_{ch}=27\%$ (at
$\theta_c = 4.7^{\circ}$, $N_c=16$, $E_c=4.97\cdot 10^{19}$~eV, with
$P_{min}=5.5\%$).}
\end{figure}

The independent data set has less
statistical power than the total data set.  If we estimate that power by
counting the number of all possible pairs (1596) among 57 events, 
then we find that
the original data set contains 27\% of those pairs, the independent
set contains 22\%, and the remaining 51\% are 
``cross'' pairs between events in the original and new data sets.  
If we are careful to avoid contamination by the original cuts, then 
we can extend the statistical power of this test by including the 
cross-correlation with the original set.  

To do this without contamination by the initial $2.5^{\circ}$ cut obtained
from the original data set, we
replace each of the three doublets in the original set with a single event
at each of their averaged positions.  We then count the number of
autocorrelation pairs in the independent set, and 
we now add the number of
cross pairs between events in the independent and original data sets.  
There is one auto pair, as before, and there are two cross pairs.
To estimate the 
chance probability, we generate Monte Carlo data sets of 27 events to
replace the independent data set, while holding the original set 
fixed.  We count what fraction of these
trials have the same or greater number of auto- and cross-correlation
pairs.  The chance probability for a total of three
or more pairs is found to be 8\%.
\footnote{
The unbiased test can be extended to include two more years of 
data since May 2000 which has been summarized on the AGASA web 
page~\cite{agasaweb}.  There are a total of 72 events above 
$E_c=4\cdot 10^{19}$\,eV through July 2002, 
which means the independent set has 42 events and roughly
double the statistical power as before.  This set adds
one new autocorrelation pair to the one already present, 
and no new cross-correlation pairs.  
Performing the same analyses as described in the text, 
the chance probability for two pairs within the independent set of 
42 events is 19\%.
The chance probability for a total of four pairs---within
the independent set, and between the independent and original sets---is 12\%.
}

We observe that if the cuts had been {\it a priori} for the first data set,
the chance probability for the three pairs among the first 30 events would
be 0.8\%.  Thus it is the clustering in the first data set which dominates
the significance for the total set, despite the fact that the first set by
itself represents only a fraction of the total number of possible pairs.
This is precisely what is to be expected when a small initial set is
used to optimize the cuts.

If we had not modified the initial data set by replacing the
doublets with their average positions, then there would have been three
cross pairs in the above test, instead of two.  The difference
is due to the events which form the AGASA triplet.  The chance
probability for four or more total pairs in this test would have been
3\% rather than 8\%.  Unfortunately, since it was these doublets in the
first set which made the $2.5^{\circ}$ cut optimal, they cannot be included
in a statistically independent test of the hypothesis.  In any case,
even this biased test confirms that the significance is dominated by the
initial data set.

\section{Conclusions}

Taken at face value, an autocorrelation scan of the published AGASA data set
finds a chance probability around $0.3\%$ for the clustering signal
claimed previously.  
At this level, the observed clustering could be a hint of real 
small-scale anisotropy, 
or it could well be a chance fluctuation in an isotropic distribution.
To investigate the possibility that it is a fluctuation, 
and that the significance of the scan is artificially high 
because it includes the data which led to the
clustering hypothesis in the first place, we test the original
claim while excluding the contribution of the original
data to the clustering signal.
First we form an independent data set using only the AGASA events 
observed after the claim.   
The cuts which were identified originally can now be applied 
{\it a priori} in an unbiased test.
To increase the statistical power of the 
test, we include cross-correlations 
with the original data.  Replacing the doublets in the 
original set with single events to keep this test independent of the 
original clustering signal, we find a chance probability of 8\%.

We conclude that the evidence for clustering in the AGASA data set is
weaker than has sometimes been claimed, and in fact is consistent with the
null hypothesis of isotropically distributed arrival directions at the 8\%
level.  This conclusion is of course not exhaustive of all the possible
anisotropies that can be studied.  For example, in~\cite{uchihori00} it was
observed (using a combination of data from AGASA and earlier cosmic ray
experiments) that the significance of the clustering increases when the
field of view is restricted to within $10^{\circ}$ of the supergalactic
plane.  The authors were careful to note that the probability ($<1\%$) of
this occurrence is {\it a posteriori}, 
and therefore not indicative of the
true significance.  Nevertheless, intriguing 
observations such as these make the need for testing
with independent data abundantly clear.  It is quite possible that with
the increased statistics and improved angular resolution of experiments
like HiRes and the Pierre Auger Array, previous claims will be decisively
tested, and compelling evidence for anisotropy may yet be found.

\ack
We are indebted to Cy Hoffman for his extremely useful suggestions 
during the writing of this paper.  We also thank Charles Jui and 
Benjamin Stokes for many stimulating discussions, and Alan Watson
for his helpful correspondence.

\end{document}